\documentclass{article}

\usepackage[utf8]{inputenc}
\usepackage[T1]{fontenc}
\usepackage{graphicx}
\usepackage{amsmath}
\usepackage{textcomp}
\usepackage{booktabs}
\usepackage{tabularx}
\usepackage[final]{microtype}
\usepackage{listings}
\usepackage{xcolor}
\usepackage{float}
\usepackage{pageslts}
\usepackage{xurl}
\usepackage{comment}
\usepackage[numbers,sort&compress]{natbib}
\usepackage{hyperref}
\usepackage{setspace}

\newif\iflongversion
\longversionfalse

\iflongversion
    \includecomment{longonly}
    \excludecomment{shortonly}
\else
    \excludecomment{longonly}
    \includecomment{shortonly}
\fi

\lstdefinestyle{json}{
    language=,
    basicstyle=\small\ttfamily,
    breaklines=true,
    breakatwhitespace=true,
    showstringspaces=false,
    stringstyle=\color{blue},
    commentstyle=\color{gray},
    keywordstyle=\color{purple},
    backgroundcolor=\color{black!5},
    frame=tb,
    framerule=0pt,
    columns=fullflexible,
    keepspaces=true,
}

\title{AI-assisted protocol information extraction for improved accuracy and efficiency in clinical trial workflows}
\author{Ramtin Babaeipour\textsuperscript{a}, Fran\c{c}ois Charest\textsuperscript{a}, Madison Wright\textsuperscript{a}\\[6pt]
\textsuperscript{a}Banting Health AI, \href{https://bantinghealth.ai}{bantinghealth.ai}\\
{\footnotesize 357 Bay St., Toronto, ON, M5H 4A6, Canada}
}
\date{\today}

\begin{document}
\label{FirstPage}

\emergencystretch=3em
\raggedbottom

\maketitle

\begin{abstract}
Increasing clinical trial protocol complexity, amendments, and challenges around knowledge management create significant burden for trial teams. Structuring protocol content into standard formats has the potential to improve efficiency, support documentation quality, and strengthen compliance. We evaluate an Artificial Intelligence (AI) system using generative LLMs with Retrieval-Augmented Generation (RAG) for automated clinical trial protocol information extraction. We compare the extraction accuracy of our clinical-trial-specific RAG process against that of publicly available (standalone) LLMs. We also assess the operational impact of AI-assistance on simulated extraction Clinical Research Coordinator (CRC) workflows.

Our RAG process shows higher extraction accuracy (89.0\%) than standalone LLMs with fine-tuned prompts (62.6\%) against expert-supported reference annotations. In simulated extraction workflows, AI-assisted tasks are completed $\geq$40\% faster, are rated as less cognitively demanding and are strongly preferred by users. While expert oversight remains essential, this suggests that AI-assisted extraction can enable protocol intelligence at scale, motivating the integration of similar methodologies into real-world clinical workflows to further validate its impact on feasibility, study start-up, and post-activation monitoring.
\end{abstract}


\onehalfspacing

\section{Introduction}\label{sec:introduction}

When properly planned and executed, clinical trials are known to be the best experimental method to evaluate the effectiveness and safety of a medical intervention. A clinical trial protocol constitutes a written agreement between investigators, research teams, participants, and the scientific community that assists communication by providing the trial's background, objectives and details about its design and organization \cite{friedman2015}. It therefore contains foundational information that teams must extract and interpret in order to ensure a consistent and compliant execution.

However, as protocol complexity increases \cite{jones2013, varse2019, getz2018}, completeness and adherence to quality guidelines of protocol documents vary \cite{gryaznov2022} and time-consuming, avoidable amendments have become more frequent \cite{getz2024}. In this context, structured data extraction from protocol documents and referencing can be time-intensive and prone to inconsistencies \cite{datta2024, kramer2025} even though it has the potential to improve downstream efficiency, support documentation quality, facilitate the ethical review process, and strengthen compliance \cite{kargren2023, georgieff2023, fda2023}. Ultimately, protocol quality and review efficiency gains would reduce burden and delays while improving evidence generation, transparency and translation to better healthcare \cite{chan2025}.

Traditionally, protocol structuring, understanding and operationalization rely on expert-driven review, iterative cross-functional clarification, and manual abstraction of key elements (e.g. endpoints, interventions, eligibility, safety, visit schedules) into spreadsheets and downstream systems (e.g. CTMS, IRB platforms, EDC). Given its highly manual execution, this process is time-consuming and introduces avoidable variability, incompleteness and inaccuracies through repeated transcription, fragmented handoffs and amendments. Comprehensive technological solutions for the extraction and mapping of unstructured protocol data to downstream systems remain nascent, with specialized tools addressing only fragmented parts of the workflow (e.g. isolated scripts, vendor-specific modules) while general-purpose ad-hoc solutions (e.g. conversational AI) typically lack integration, repeatability, performance and compliance.

However, LLMs are AI systems capable of distilling complex, unstructured information into key data elements and summaries. They can serve as an assistive layer in established workflows, generating standardized first versions of data and documents that can be submitted to expert verification. By reducing time spent on routine structuring and document navigation, such systems can help teams focus on higher-leverage activities (e.g. adjudication of ambiguous cases, quality oversight) while improving consistency and auditability of protocol-derived data.

Beyond base information extraction, they provide automated reasoning and content generation capabilities. The latter have the potential to accelerate protocol document standardization through automated document authoring (e.g. \cite{maleki2024}), but their more immediate application lies in extracting structured information from existing unstructured protocol documents and providing initial automated analyses. While \cite{babaeipour2026} will go beyond information extraction, evaluating the RAG methodology for producing protocol complexity estimations under publicly available frameworks (e.g. scores and rationales over lists of complexity domains), this first paper focuses on information extraction.

We design, implement and evaluate a novel clinical-trial-specific RAG system with a SoE-specific methodology for extensive automated protocol information extraction. It combines domain-specific RAG for text-based representations with specialized vision-based methods for tabular SoE data, enabling comprehensive extraction across diverse protocol structures. This addresses the foundational limitations of both standalone LLMs and general-purpose extraction approaches, extending the reach and scale of extraction procedures found in the current literature.

We empirically compare our RAG approach against standalone LLMs across 23 publicly available protocols spanning multiple therapeutic areas, demonstrating improved accuracy particularly for complex, scattered information. Through a controlled experiment with 13 CRCs, we assess information extraction accuracy as well as real-world operational impact, measuring time savings, cognitive load, and user preferences. In order to achieve a robust evaluation methodology, we developed an LLM-assisted evaluation and annotation adjudication framework that enables scalable, consistent assessment across hundreds of semi-structured data fields (see table \ref{table:tab1}).

\begin{table}[ht!]\label{table:tab1}
    \centering
    \caption{Statement of significance}
    \begin{tabular}{p{0.27\linewidth} p{0.68\linewidth}}
        \toprule
        \textbf{Problem or issue} & Clinical trial protocols are long, heterogeneous PDFs; extracting accurate, auditable, structured operational data (e.g., endpoints, interventions, SoE) is time-consuming, inconsistent, and difficult to scale. \\
        \midrule
        \textbf{What is already known} & LLMs are used to extract eligibility and schedules, RAG improves precision by retrieving relevant context and vision-enhanced LLMs can interpret tables. \\
        \midrule
        \textbf{What this paper adds} & A clinical-trial-specific RAG pipeline plus a SoE table extraction approach that produces traceable, semi-structured outputs, an empirical comparison vs. standalone LLMs, a CRC controlled experiment showing faster, more accurate extraction, and a LLM-based evaluation framework for scalable, consistent performance assessment. \\
        \midrule
        \textbf{Who would benefit} & Site feasibility, start-up, operations and monitoring, site IT (CTMS/EDC, downstream automation), CROs and sponsors. \\
        \bottomrule
    \end{tabular}
\end{table}

\section{Related work}\label{sec:related_work}

The current literature shows LLMs being used to extract from study documents certain study design features such as eligibility criteria \cite{datta2024, liu2021} and study schedule of events \cite{kramer2025, snorkel2022}. Recent examples feature the extraction of more general semi-structured information from unstructured oncology medical records \cite{wiest2025}. This paper presents a more comprehensive extraction approach.

When extracting information from unstructured documents, a direct approach involves prompting a LLM with detailed instructions and providing entire documents as part of its context (standalone LLM). While recent LLMs allow for very large context windows, this approach has theoretical limitations. Among them:
\begin{itemize}
    \item \textbf{Context window limits}: These LLMs still can only work with limited document lengths, and protocol documents may exceed their input token limits \cite{hosseini2024},
    \item \textbf{Context window spread}: they may not consistently identify and extract all relevant information, especially when details are scattered across different sections of lengthy protocols \cite{liu2024},
    \item \textbf{Query number tradeoff}: on one hand, using a small number of prompts to extract hundreds of independent, individual data elements may lead to suboptimal performance, while on the other hand, using many prompts with such large contexts increases cost and time to completion \cite{lewis2020},
    \item \textbf{Lack of element-specific context and referencing}: each data element extracted may require specialized context and prompting, knowledge of clinical research terminology and output requirements \cite{rajpurkar2022}. Practitioners may also need to reference specific protocol sections for auditability and traceability, which standalone LLMs may not provide in a natural way.
\end{itemize}

Retrieval-Augmented Generation (RAG) \cite{lewis2020} addresses these limitations by combining the general knowledge encapsulated in LLMs with element-specific information retrieval queries, context and generation prompts.

Since the Schedule of Events (SoE) defines study timing and procedures on which numerous downstream processes depend, it is crucial for operational execution \cite{jscdm2025}. Clinical trial protocols very frequently use table formatting to represent information, most notably the SoE. Furthermore, SoE often involves multi-page spans, intricate cell merging, and hierarchical visit structures encoded through visual layout.

Current PDF extraction methods often struggle with those particular challenges, from markup conversion \cite{ferres2018} (cumbersome and labor-intensive) and image-based recognition \cite{zhong2020} (dominant approach, but heavily dependent on trained models) to text and metadata direct extraction (hard to generalize). Because SoE formatting varies widely across protocols, traditional rule-based or metadata-dependent approaches are ruled out, failing to reliably capture the hierarchical relationships. Instead, we address SoE extraction through a specialized two-stage approach involving table detection on protocol pages followed by vision-based multimodal generation on those pages for information extraction. This approach seems sufficient to achieve similar performance levels on SoE than on other information categories.

Another challenge requiring specific consideration is the quality evaluation of semi-structured data outputs. Recent work has demonstrated the effectiveness of using LLMs as evaluators, often termed ``LLM-as-a-judge'', for assessing the quality of AI-generated content. A comprehensive survey shows that in many settings LLM-based evaluation correlates well with human judgments \cite{gu2024}. In healthcare specifically, it has been demonstrated that gpt-4o can effectively automate the evaluation of AI-generated clinical text, achieving strong agreement with expert clinicians while significantly reducing evaluation time and cost \cite{croxford2025}. This approach is particularly valuable in clinical research where human expert evaluation is resource-intensive, yet maintaining quality standards is critical. Building on this foundation, we employ an LLM-based evaluation framework to assess the accuracy and completeness of protocol abstractions in our study.

In addition to evaluation being subject to variations in output phrasing, there is also a subjective nature to what constitutes the ground truth, as experts sometimes differ on what constitutes necessary and sufficient extracted information. Moreover, given the document density and breadth of the annotation task, even expert reviewers are prone to incomplete data capture. This means that a small group of expert annotators cannot realistically guarantee a truly exhaustive ground truth. While increasing the number of human reviewers could improve completeness, such a redundant process is prohibitively labor-intensive and fundamentally unscalable. In many recent comparable works (\cite{yuan2025}, \cite{wang2024}, \cite{thomas2024}), researchers introduce LLMs in their annotation process, both on generating final adjudication candidates and in adjudicating those outputs. Similarly, we designed a LLM-assisted annotation process that presents a hybrid human--AI collaboration to an independent LLM-based adjudication layer whose output is human-reviewed on low-confidence cases and quality-controlled on a randomly sampled subset.

\section{Methods}

\subsection{Protocol documents selection}\label{sec:protocol_selection}

Starting from all the studies listed in clinicaltrials.gov \cite{ctgov} at the end of March 2025 having an identifiable protocol document, we select interventional drug studies with treatment as primary purpose, conducted in Canada or the United States, from which we randomly sample studies: 9 from oncology, 7 from cardiovascular, and 7 from other therapeutic areas (see Listing \ref{lst:filtering} for details).

For those studies, a human data annotation expert manually creates a semi-structured dataset from the data models described in Section~\ref{sec:extracted_models}.

\subsection{Extracted semi-structured data models}\label{sec:extracted_models}

In order to structure and evaluate the information extraction accuracy of different methodologies, we define a set of semi-structured data models representing key data elements to be extracted from clinical trial protocols. Those data models are designed to capture essential protocol information in a standardized, json-representable format that simplifies data handling, comparisons and evaluation.

We classify the information to be extracted into six broad categories: general information, inclusion/exclusion criteria, adverse event definitions, intervention, site requirements and schedule of events. Each category is subdivided into smaller data elements representing the relevant (semi-)structured output to be extracted by the RAG invocations, and also serve as a basis for performance evaluation. The list of those data elements is provided in the following subsections and Table~\ref{tab:component_weights}.

\subsubsection{General information}

Elements from this category include general study information (e.g. NCT ID, protocol version, title, sponsor, phase, therapeutic area, disease/condition, allocation, masking, estimated duration) as well as primary and secondary objectives and endpoints (see Listing \ref{lst:gen_info} for an example).

\subsubsection{Interventions}

Elements from this category include trial arms (name, description) and intervention details (type, name, dosage, schedule) as well as treatment-level information (product name, dose, administration, restrictions, modifications) (see Listing \ref{lst:intervention} for an example).

\subsubsection{Schedule of events}

Elements from this category include visit number, visit time, and procedures to be performed at each visit (see Listing \ref{lst:soe} for an example).

\subsubsection{Inclusion/exclusion criteria}

Elements from this category include inclusion criteria and exclusion criteria (see Listing \ref{lst:ie} for an example).

\subsubsection{Adverse event definitions}

Elements from this category include adverse event (AE) and serious adverse event (SAE) definitions, severity grading, relationship to study treatment, reporting requirements (timeframes, data collection, contacts), safety monitoring and management (plan, discontinuation criteria, emergency procedures), and specific AE information (expected AEs, potential risks, concomitant medication restrictions, special population considerations) (see Listing \ref{lst:ae_definitions} for an example).

\subsubsection{Site requirements}

Elements from this category include site equipment, certifications, sample handling requirements as well as investigational product (IP) storage conditions (see Listing \ref{lst:site_req} for an example).

\subsection{Data extraction approaches}

We use the following three data extraction approaches.

\subsubsection{RAG extraction}\label{sec:rag_extraction}

\begin{figure}[ht!]
    \centering
    \makebox[\textwidth][c]{\includegraphics[width=1.2\textwidth]{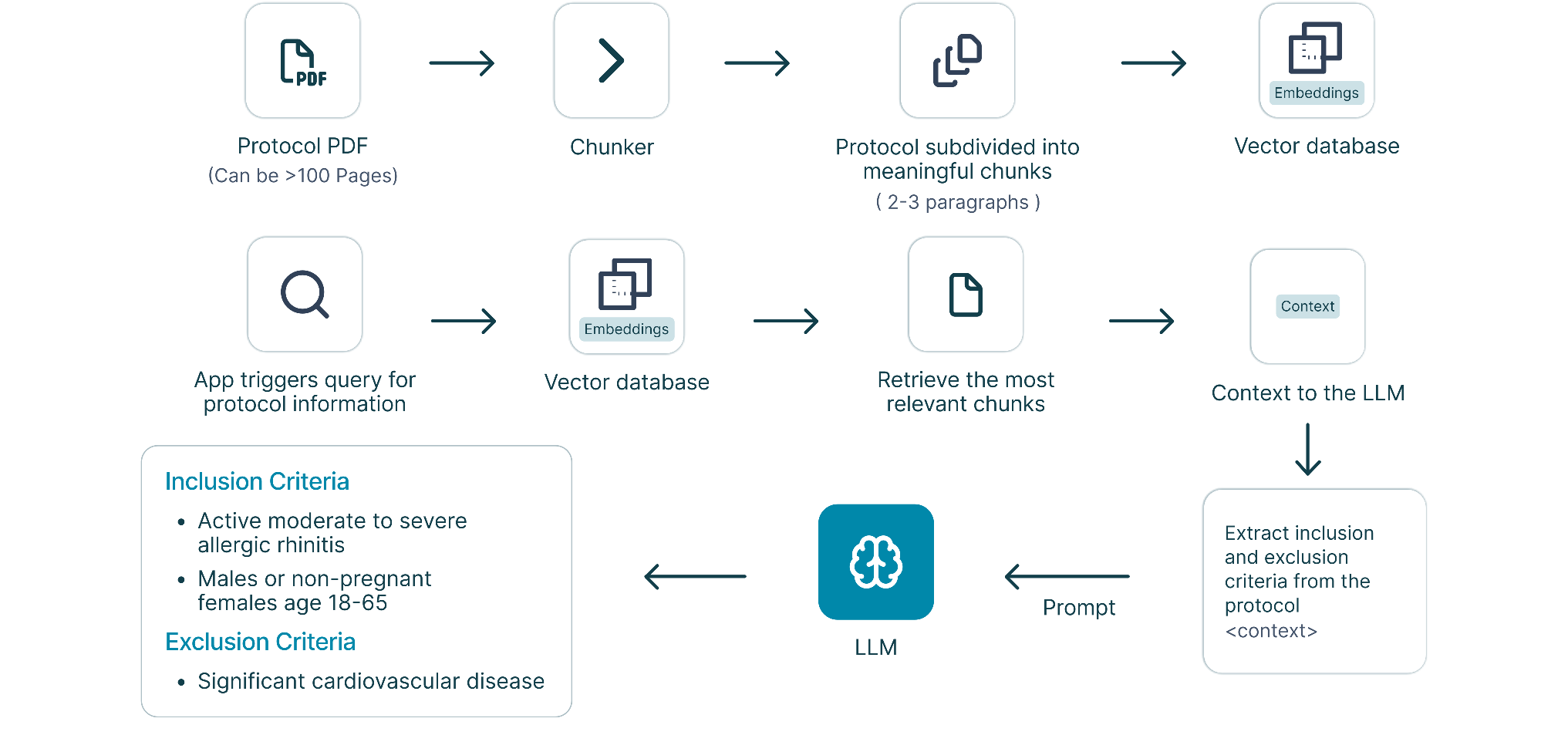}}
    \caption{RAG process for clinical protocol information extraction. The RAG system first processes protocol Portable Document Format (PDF) files subdividing them into meaningful chunks and storing them in a vector database with semantic embeddings. When users query for specific protocol information (e.g., inclusion/exclusion criteria), the system retrieves the most relevant chunks and provides them as context to a LLM along with specialized instructions. The LLM then generates a (semi-)structured, standardized output that research teams can readily use for protocol review and study operations.}
    \label{fig:rag_process}
\end{figure}

Our clinical-trial-specific RAG process works in three key steps, illustrated in Figure~\ref{fig:rag_process}:

\begin{enumerate}
    \item \textbf{Document processing}: Protocol documents are subdivided into manageable chunks of information, vectorized using a distinct embedding model, and stored in a searchable database preserving context and relationships between sections.
    \item \textbf{Custom retrieval queries}: When seeking specific information (e.g. primary endpoints, inclusion criteria), the system uses tailored queries to identify and retrieve the most relevant chunks from the protocol document. Referencing those chunks also allows for direct auditability and traceability.
    \item \textbf{Structured information generation}: A generation LLM is presented with the retrieved information as context along with custom instructions (prompts) designed specifically for clinical research data extraction, and required to structure the information according to the required formats (json schemas).
\end{enumerate}

This approach allows the system to work with protocol documents of any length, focusing on relevant sections for each element-specific query, to leverage clinical research expertise embedded in the retrieval and prompting strategies, and to provide transparent referencing for human approval and process auditability supporting regulatory electronic records compliance. Moreover, since data privacy and security are paramount in clinical research, our RAG system is designed to operate within secure, compliant environments, ensuring that all information involved remains private throughout data processing (e.g. it cannot be used by third parties for LLM training).

This RAG approach features modular retrieval and prompting strategies: each data element or small group of closely related data elements (e.g. `protocol id', `protocol date' and `protocol version') is the target of a dedicated RAG chain call, and the LLM API is then provided with only the relevant chunks retrieved from the protocol document as the context alongside a element-tailored prompt.

Moreover, in this setup,
\begin{itemize}
    \item we use a vector embedding model (Cohere English v4 Embeddings (RAG~1) and Cohere English v3 Embeddings \cite{cohereembeddings} (RAG~2)) to vectorize the protocol document chunks,
    \item we employ methods of context aware chunking and chunk reunion based on section headers in addition to document ordering in the retrieval steps as described below,
    \item we use, among others, gpt-4o and Claude generation LLMs,
    \item we source standard methods and orchestrate our RAG process through the LangChain framework \cite{langchain2022} for improved abstraction, observability and portability.
\end{itemize}

The controlled experiment (Section~\ref{sec:controlled_experiment}) uses an earlier RAG configuration with gpt-4o~mini as the generation model used at the time of data collection. This configuration is reported as RAG~2 in the comparative evaluation (Section~\ref{sec:rag_vs_standalone}), ensuring that the controlled experiment remains internally consistent with this RAG configuration. Replicating the controlled experiment (which involves consenting and coordinating 13 CRCs across a randomized, controlled design) solely to align with the later Claude Sonnet~4.5 configuration (RAG~1) is excluded from the scope of this study. The RAG~1 results instead serve to independently demonstrate the system's improvement trajectory and upper-bound performance potential, separate from the CRC workflow evaluation.

Recent research in RAG has demonstrated that chunking strategy significantly impacts retrieval quality and generation accuracy \cite{wang2024searching}. For clinical trial protocols that have hierarchical organization and cross-referenced sections, naive chunking approaches often break apart related information or mix unrelated content together. The main challenge is maintaining a balance between providing enough context and ensuring relevance, as large chunks include unrelated information while small ones lose essential context \cite{li2025enhancing}.

To address these challenges, we implement three complementary techniques. First, context-aware chunking identifies and preserves hierarchical document structure (section headers, subsections, numbered lists), ensuring chunks retain semantic coherence and metadata about its position in a hierarchy. Second, with automatic chunk reunion, split portions of oversized sections are programmatically reunited during retrieval. Third, document ordering sorts results by their original sequence, maintaining the logical flow and cross-referential integrity. These methods combined ensure that the LLM receives contextually complete, structurally coherent, and logically ordered information for each query.

\subsubsection{Standalone LLM extraction}

We use a standalone LLM approach as a baseline for comparison. We use a large generation LLM (gpt-4o \cite{gpt4o}), with similarly engineered, but larger, less specific prompts encompassing the elements from the categories from Section~\ref{sec:extracted_models}. Therefore, for each data element category, we have a single prompt that asks the LLM to extract all elements from that category, and the LLM API is provided with the entire protocol document text in its context.

Note that ignoring scalability, security and compliance considerations, this benchmark approach is reminiscent of a research team member uploading the study protocol to the equivalent publicly available general-purpose AI-assistance service, and submitting carefully engineered prompts for the six data categories.

\subsubsection{Schedule of Events table detection and extraction}

SoE information presents unique extraction challenges due to its typical presentation in complex tabular formats with intricate spatial relationships between visits, procedures, and timepoints. We address this through a specialized two-stage approach: table detection (transformer-based) followed by multimodal LLM information extraction (native vision capability). This decomposition is deemed necessary because clinical protocol tables often span multiple pages, exhibit diverse layouts across protocols, and embed critical scheduling information within dense visual structures that resist purely text-based parsing methods.

For table detection, we employ the Microsoft Table Transformer \cite{smock2022}, an object detection model trained on tables from scientific articles. Multimodal LLMs have demonstrated effective visual cognition capabilities \cite{schulze2025}. We then submit the identified table pages (converted to images) to a multimodal LLM with SoE-specific prompting instructions, attempting to maintain context across multi-page tables and tie procedures to visits. This multimodal approach is effective for SoE extraction, mostly preserving the spatial layout and visual cues that encode hierarchical relationships in table structures while it is regularly lost or inaccurate in text-only representations. In our more recent configuration (RAG~1), we use Claude Opus~4.5 while our earlier configuration (RAG~2) uses gpt-4o. As in Section~\ref{sec:rag_extraction}, the controlled experiment uses the RAG~2 configuration (gpt-4o for SoE) while RAG~1 is used to demonstrate the performance gains achievable with more recent multimodal models independently of the CRC workflow evaluation.

\subsection{Evaluation}

Since comparing hundreds of semi-structured fields from each protocol to their ground truths is resource-intensive, non-scalable, error-prone and involves hard-to-manage variability across humans, we adopt scripted procedures leveraging LLMs (gpt-4o) to compare the extracted semi-structured content to its ground truth.

Evaluations are done at the same level of granularity as the RAG chain calls, so for each data extraction RAG chain call, the output json object is evaluated independently of others. Therefore, the majority of the jsons evaluated contain only a few data elements, if not just one. This allows for more robust and explainable evaluation and enables comparisons on constrained data elements.

The prompting used for evaluation of those components is general, focused on judging the completeness and precision of the data instead of the style or structure. For instance, we instruct the LLM-as-judge to not be sensitive to the use of abbreviations, to consider ``age $\geq$ 60'' as equivalent to ``Age 60 or older'', ``Merck \& Co., Inc.'' as equivalent to ``Merck'', ``10.000'' as equivalent to ``10'', ``10,000'' as equivalent to ``10\,000'', etc. Moreover, we instruct the LLM-as-judge to not penalize for extra information that does not contradict the ground truth, misrepresent information or introduce safety risks. This is intended to account for different levels of verbosity and detail in the extracted information across humans and LLMs, but ensuring that potential LLM hallucinations are penalized.

A score from 0 to 5 (values 0, 1, 2, 3, 3.5, 4, 4.5 and 5) is assigned based on how closely the submitted json reflects the entire information from the ground truth json, with 5 meaning that the submitted json recovers 100\% of the ground truth, 4.5 meaning that it recovers at least 90\% of the ground truth json to a human reader, and 0 meaning that it recovers less than 20\% of the ground truth.

In order to obtain a score for a given data element category (e.g. general information, interventions, etc.), predefined weights are used in a weighted average of the scores of all the individual jsons evaluated for that category. This allows us to compare the performance of different extraction approaches on each category while ensuring that, for example, primary endpoints and objectives are weighted higher than their secondary counterparts (e.g. in the general information category, \texttt{primary\_objectives} have a weight of 0.25 while \texttt{secondary\_objectives} have a weight of 0.02). See Table~\ref{tab:component_weights} in Appendix~\ref{sec:annexes} for the complete list, and Section~\ref{sec:evaluation_examples_annex} for evaluation examples.

To validate the reliability of the LLM-as-judge framework, we assembled a benchmark dataset of 95 scored pairs of json elements spanning diverse data element categories and protocols, independently scored by both a human expert and the LLM judge. Following recent approaches to empirical AI evaluation reliability assessment \cite{gu2024, croxford2025, paperbench2025}, we report agreement statistics in Table~\ref{tab:judge_validation}. The LLM judge achieves strong correlation (Pearson $r = 0.923$, Spearman $\rho = 0.942$, both $p < 0.001$) and weighted agreement (linear $\kappa = 0.699$, quadratic $\kappa = 0.877$) with human scores. A small negative bias of $-0.47$ points (on a 0--5 scale) indicates the LLM judge tends to score slightly lower than the human evaluator, explaining most of their disagreements.

\begin{table}[ht!]
    \centering
    \caption{LLM-as-judge vs. human expert agreement statistics ($N = 95$ scored pairs).}
    \label{tab:judge_validation}
    \begin{tabular}{lc}
        \toprule
        \textbf{Metric} & \textbf{Value} \\
        \midrule
        Mean Absolute Error (MAE)       & 0.553 \\
        Bias (AI $-$ Human)             & $-0.468$ \\
        Pearson $r$                     & 0.923\ \ ($p < 0.001$) \\
        Spearman $\rho$                 & 0.942\ \ ($p < 0.001$) \\
        Weighted $\kappa$ (linear)      & 0.699 \\
        Weighted $\kappa$ (quadratic)   & 0.877 \\
        Exact agreement ($\pm 0$)       & 43.2\% \\
        Near agreement ($\pm 0.5$)      & 64.2\% \\
        \bottomrule
    \end{tabular}
\end{table}

This evaluation methodology is used to compare both the standalone LLM to the RAG approach and the AI-assisted tasks to the manual tasks from Section~\ref{sec:controlled_experiment}.

\subsection{Data annotation}\label{sec:data_annotation_approach}

To achieve the scale and breadth required to allow for proper performance assessment, we draw upon recent methodologies that integrate LLMs into the annotation process to augment human capabilities. For example, recent work demonstrated the efficacy of ``multi-LLM consensus'' combined with human review to achieve scalable content annotation \cite{yuan2025}. Similarly, human-LLM collaboration, specifically where humans verify LLM-generated labels rather than annotating from scratch, significantly improves efficiency without sacrificing accuracy \cite{wang2024}. Recent examples show an application where LLM labeling can match or exceed human agreement with ``gold standard'' ground truth \cite{thomas2024}. These all support the idea that LLMs can be used within high-quality annotation and adjudication procedures, often being comparable to trained humans, especially when guided with optimized prompts and calibration.

Guided by these findings, we design a hybrid human--AI set of annotations that is submitted to an independent LLM-based adjudication layer. Low-confidence adjudications are sent to human experts for final review, as well as a random sample of data elements used as quality control the LLM adjudication process itself. This hybrid approach aims to balance the scalability and speed of LLMs with the domain expertise and critical judgment of human reviewers, ultimately yielding a high-quality ground truth dataset.

\subsubsection{Annotation candidates}

First, we use a human-annotated full dataset from the 23 protocols (covering all data elements of all the data categories). We generate two more candidate annotation datasets leveraging different generative models, one from OpenAI (gpt-4o), the other one from Anthropic (Claude 3).

\subsubsection{LLM-based adjudication of annotation candidates}

We employ a state-of-the-art LLM (gpt-5) to serve as the primary adjudication engine. This model acts as an independent evaluator, synthesizing information from:
\begin{itemize}
    \item the original protocol document (full PDF text provided),
    \item the extraction instructions used by RAG chains for each data element and
    \item the 3 candidate annotations, presented in a random order, without reference to their provenance.
\end{itemize}

The instructions provided to the LLM adjudicator emphasize generating
\begin{itemize}
    \item a final annotation in strict adherence to the information found in the protocol and the predefined instructions and schema for each data element. This final json output can be either of the submitted candidate annotations, or a new annotation synthesized by the LLM adjudicator.
    \item a justification for each adjudication decision with clear rationales grounded in specific sections of the protocol text, and
    \item a confidence score (0--100) reflecting the model's certainty in its adjudication decision.
\end{itemize}

\subsubsection{Human-in-the-loop validation}

To ensure the reliability of this ground truth, we implement a twofold human expert review strategy involving two independent expert adjudicators.

Uncertainty sampling is used to mitigate errors in complex cases, utilizing the confidence score generated by the LLM as a filter for selection. We identify the 40 data elements with the lowest confidence scores across the dataset. These flagged elements are split between the two human adjudicators for manual review. We observe that human adjudication confirms the LLM's decision in 87\% ($N=35$) of these low-confidence cases, suggesting that the model remains robust, aligned with its confidence scores, in areas of lower certainty.

Quality control leveraging random sampling is used to establish a baseline for the overall adjudication quality. A subset of 24 randomly selected data elements (exclusive of the low-confidence set) is reviewed by both of the two human adjudicators. In this random subset, human adjudication confirms the LLM's output in 100\% ($N=24$) of cases.

Based on both of these high concordance rates, we estimate that this hybrid annotation methodology yields a ground truth accuracy comparable to fully manual double-blind annotation.

\subsection{Controlled experiment}\label{sec:controlled_experiment}

The main objective of this controlled experiment is to compare the effectiveness and accuracy of the AI-assisted approach to manual extraction by CRCs for one given RAG configuration.

\subsubsection{Experiment design}

In this experiment, data extraction tasks based on the specifications from Section~\ref{sec:extracted_models} are randomly assigned to a group of CRCs to simulate a clinical trial site workflow. A total of 13 CRCs with varying levels of experience (2 with $<$1 year, 9 with 1--3 years, 1 with 4--6 years of experience) are selected; they provide consent and receive (video and textual) instructions as well as examples of the expected output formats for each data element.

In an online survey platform, after optional demographic/experience/interest questions (gender, age group, experience as CRC, degree of comfort with new software or technologies, enthusiasm for new software or technologies), participants are presented with two tasks based on two different protocols chosen randomly from the 23 available (Section~\ref{sec:protocol_selection}). For each task, the CRC extracts 4 randomly selected data element categories from the 6 described in Section~\ref{sec:extracted_models}, and enters the values into provided spreadsheets where each data element value has its own cell.

In one of the two tasks, the CRC is provided with the RAG process AI-extracted data pre-filled in the spreadsheet, and is asked to review, complete and correct it as needed (AI-assisted task). In the other task, the CRC fills in the spreadsheet manually from scratch (unassisted task). The order of tasks (unassisted vs. AI-assisted), as in the case of protocols and data element categories, the order in which they are presented is randomized to avoid bias.

It is important to note that the AI-extracted data pre-filled in the spreadsheets is generated using a RAG configuration having gpt-4o-mini as the generation model available at the time of the experiment, not the later Claude Sonnet~4.5-based system that achieves the best evaluations in Section~\ref{sec:rag_vs_standalone}.

After each task, CRCs self-report time-to-completion and two ratings evaluating how mentally demanding the task is and how much effort they have to invest. After completing both tasks, they are asked 5 questions to assess their overall satisfaction and experience with the AI assistance (see Table~\ref{tab:user_satisfaction}).

\section{Results}

\subsection{Specialized RAG vs. standalone LLM performance}\label{sec:rag_vs_standalone}

Overall, on information extraction into the semi-structured data models, in the context of gpt-4o, our RAG process achieves an average accuracy of 69.6\% across the six categories from Section~\ref{sec:extracted_models}, comparing favorably to publicly available LLMs with fine-tuned prompting that obtains an average accuracy of 62.6\%. However, later configurations involving Claude Sonnet~4.5 and Claude Opus~4.5 and further optimizations (e.g. prompting, retrieval) achieve an average accuracy of 89.0\%.

\setlength{\tabcolsep}{4pt}
\begin{table}[ht!]
    \centering
    \caption{Category-wise RAG vs. standalone LLM score comparison}
    \label{tab:rag_vs_norag}
    \begin{tabular}{lccccc}
        \toprule
        \textbf{Category}
        & \textbf{RAG 1}
        & \textbf{RAG 2}
        & \textbf{Standalone}
        & \textbf{$N$}
        & \textbf{$N$} \\
        &
        &
        &
        & \textbf{(RAGs)}
        & \textbf{(Standalone)} \\
        \midrule
        Adverse events      & \textbf{92.3\%} & 78.0\% & 59.5\% & 23 & 22 \\
        General information & \textbf{83.5\%} & 76.1\% & 78.6\% & 23 & 22 \\
        Inclusion/exclusion & \textbf{91.7\%} & 77.6\% & 90.6\% & 23 & 22 \\
        Interventions       & \textbf{92.8\%} & 68.6\% & 53.1\% & 23 & 22 \\
        Site information    & \textbf{81.2\%} & 46.2\% & 38.6\% & 23 & 22 \\
        Schedule of events  & \textbf{92.7\%} & 71.3\% & 55.2\% & 22 & 21 \\
        \midrule
        \textbf{Average (weighted)}
                           & \textbf{89.0\%} & 69.6\% & 62.6\% & -- & -- \\
        \bottomrule
    \end{tabular}

    \vspace{2mm}
    \small
    \textit{Notes:}
    RAG~1: Claude Sonnet~4.5 for generation LLM, Claude Opus~4.5 for SoE, and Cohere English v4 for embeddings; RAG~2: gpt-4o~mini for generation LLM, gpt-4o for SoE, and Cohere English v3 for embeddings (used in Section~\ref{sec:controlled_experiment}); Standalone: gpt-4o.
    $N$ denotes the number of evaluated protocols per category.
\end{table}

Of the 23 selected protocols, 1 exceeds the standalone LLM's context limit and is excluded from standalone analysis, while another lacks a clear schedule of events and is excluded from that category.

The results demonstrate distinct performance patterns across categories. For general information and inclusion/exclusion, both approaches achieve relatively close performance, considering that the lower RAG~2 inclusion/exclusion score relative to RAG~1 (77.6\% vs.\ 91.7\%) is partly attributable to the smaller chunk size limit of Cohere English v3 embeddings used in RAG~2. Eligibility criteria are often long, enumerated lists spanning multiple protocol pages, making them more prone to fragmentation across retrieval chunks in ways that lose cross-criterion context. The larger chunk capacity of Cohere English v4 in RAG~1 mitigates this fragmentation.

However, RAG's specialized and focused approach substantially surpasses standalone LLM performance on more complex extraction tasks like adverse events, site information and interventions. Since information for these categories is typically spread across different sections of the protocol, the specialized RAG system's ability to retrieve and synthesize relevant content from multiple locations provides a significant advantage. Moreover, our specialized SoE extraction approach demonstrates our approach's advantage.

An important consideration when interpreting these category-specific results is that the subjectivity level of ground truth extractions varies across element types. Certain categories, such as inclusion and exclusion criteria, general protocol information, and schedule of events have relatively low-variability representations (e.g. an exact copy of a protocol excerpt). Other categories like adverse event information, intervention details, and site-specific information can be validly represented in multiple ways depending on the level of granularity, organization, and relevance chosen in the annotations. This phenomenon of ground truth variability in specialized domains has been documented in semantics data evaluation \cite{aroyo2015}, notably in domains where expert-authored references cannot exhaustively capture all valid authorities and lines of reasoning \cite{yu2025dece}. In such context, evaluation metrics may underestimate extraction performance when responses use valid but non-standard representations or do not capture all valid nuances of various relevance. Hence, some share of performance differences may be explained by ground truth variability rather than solely extraction capability, especially since the annotation process (Section~\ref{sec:data_annotation_approach}) is aimed at capturing a very high level of detail.

\subsubsection{Observations and examples}

Across the various data element categories, several patterns emerged that highlight strengths of the RAG approach over the standalone LLM method.

\paragraph{Context confusion mitigation}

The non-RAG approach frequently suffers from context confusion when processing lengthy protocols. In one protocol (NCT01742117), when extracting adverse event definitions, the non-RAG system incorrectly focuses on bleeding event classifications:

\begin{quote}
\textbf{Standalone LLM output:} ``All bleed events that require some form of intervention (e.g. doctor's visit, ER visit, hospitalization) will be captured, recorded, and classified as major, minor, or minimal.''
\end{quote}

In contrast, the RAG system accurately retrieves and extracts the actual adverse event definition:

\begin{quote}
\textbf{RAG output:} ``An adverse event (AE) in the context of this study is defined as any problem or event which, in the opinion of the local investigator, was unanticipated, places subjects or others at a greater risk of harm than was previously known or recognized, and was possibly related to the research procedures...''
\end{quote}

\paragraph{Comprehensive information extraction}

The RAG system's targeted retrieval capability enables more thorough extraction of complex, structured information. When identifying expected adverse events (in NCT02971956), the non-RAG approach provides only generic statements:
\begin{quote}
\textbf{Standalone LLM output:} ``Adverse events (both non-serious and serious) associated with pembrolizumab exposure may represent an immunologic etiology.''
\end{quote}
The RAG system, however, retrieves and structures comprehensive lists of specific expected events:
\begin{quote}
\textbf{RAG output:} ``[ ``Pneumonitis'', ``Colitis'', ``Endocrine disorders (e.g., Adrenal Insufficiency, Hyperthyroidism, Hypophysitis, Hypopituitarism, Hypothyroidism, Thyroid disorder, Thyroiditis)'', ``Hematologic disorders (e.g., Thrombotic Thrombocytopenic Purpura, Autoimmune hemolytic anemia, Aplastic anemia...)'', \ldots ]''
\end{quote}

The non-RAG system's performance declines notably in protocols exceeding typical token limits, where relevant information becomes diluted within extensive background content, where critical details are often scattered across multiple protocol sections. The RAG system's ability to retrieve targeted sections addressing specific queries allows it to maintain better performance regardless of overall document length.

\subsection{Controlled experiment results, AI-assisted vs. manual data extraction}\label{sec:controlled_experiment_results}

Each of the 13 CRCs is assigned two protocols: one with AI-assisted extraction (spreadsheet pre-filled with AI-generated content for review and correction) and one with manual extraction (starting from an empty spreadsheet). For each protocol, CRCs extract four randomly selected data element categories from the six described in Section~\ref{sec:extracted_models}.

Under the original design, this yields 104 completed extraction tasks (13 CRCs $\times$ 2 protocols $\times$ 4 categories). However, the final dataset comprises 97 tasks due to two factors: (1) four CRCs complete only three of the four assigned categories for one of their protocols (4 tasks), and (2) one protocol lacks a schedule of events table, necessitating the exclusion of schedule of events extraction tasks for the three CRCs assigned to that protocol (3 tasks), see Table~\ref{tab:subtask_distribution}.

\begin{table}[ht!]
    \centering
    \caption{Distribution of data extraction subtasks by type}
    \label{tab:subtask_distribution}
    \begin{tabular}{lccc}
        \toprule
        \textbf{Category} & \textbf{AI-assisted} & \textbf{Manual} & \textbf{Total} \\
        \midrule
        Adverse events        & 10  & 9  & 19 \\
        General information   & 8   & 11 & 19 \\
        Inclusion/exclusion   & 7   & 5  & 12 \\
        Interventions         & 11  & 10 & 21 \\
        Sites                 & 7   & 7  & 14 \\
        Schedule of events    & 7   & 5  & 12 \\
        \midrule
        \textbf{Total} & 50 & 47 & 97 \\
        \bottomrule
    \end{tabular}
\end{table}

\subsubsection{Efficiency of data extraction}

\begin{figure}[ht!]
    \centering
    \includegraphics[width=0.8\textwidth]{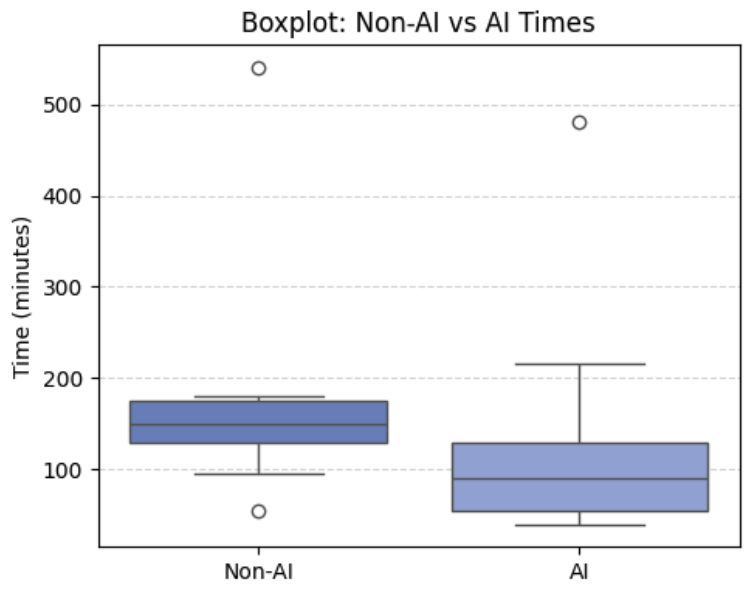}
    \caption{Comparison of task completion times between non-AI and AI-assisted protocol abstraction tasks (AI-assisted: RAG~2, gpt-4o-mini for generation, gpt-4o for SoE). Box plot shows median completion time, interquartile range, and outliers (circles) for both conditions. Non-AI tasks required substantially longer completion times compared to AI-assisted tasks.}
    \label{fig:completion_times}
\end{figure}

Overall, AI-assisted tasks are completed faster than manual tasks (see Figure~\ref{fig:completion_times}). A one-sided paired (within subject, repeated measurement) t-test shows that the difference is statistically significant ($p=0.008$), see Table~\ref{tab:completion_times}.

\begin{table}[ht!]
    \centering
    \caption{Comparison of task completion times between AI-assisted and manual protocol abstraction tasks. AI-assisted using RAG~2 (gpt-4o-mini for generation, gpt-4o for SoE extraction).}
    \label{tab:completion_times}
    \begin{tabular}{lcc}
        \toprule
        \textbf{Condition} & \textbf{Median time (min)} & \textbf{Mean time (min)} \\
        \midrule
        AI-assisted   & 90.00  & 120.67 \\
        Manual        & 150.00 & 167.67 \\
        \bottomrule
    \end{tabular}
\end{table}

This translates to an average time reduction of 47 min per protocol and a median reduction of 60 min (40\% improvement). The effect size indicates a medium-to-large practical impact on workflow efficiency.

\subsubsection{Quality of data extraction}

AI assistance significantly improves the accuracy and completeness of protocol abstractions. The item-weighted average score is substantially higher for AI-assisted tasks compared to non-AI tasks (see Table~\ref{tab:quality_scores} for reporting average per item for the two tasks, Welch's independent groups t-test: $t=2.62$, $p<0.05$). This represents a medium effect size (Hedges' $g=0.53$), indicating practical significance beyond statistical significance.

\begin{table}[ht!]
    \centering
    \caption{Comparison of evaluation scores by item. AI-assisted using RAG~2 (gpt-4o-mini for generation, gpt-4o for SoE extraction).}
    \label{tab:quality_scores}
    \begin{tabular}{lcccccc}
        \toprule
        & \multicolumn{3}{c}{\textbf{AI assisted}} & \multicolumn{3}{c}{\textbf{Unassisted}} \\
        \cmidrule(r){2-4} \cmidrule(l){5-7}
        \textbf{Category} & \textbf{Score} & \textbf{Score (/100)} & \textbf{Count} & \textbf{Score} & \textbf{Score (/100)} & \textbf{Count} \\
        \midrule
        ae    & 3.8155 & 76.31 & 10 & 2.2751 & 45.50 & 9 \\
        gen   & 4.0062 & 80.12 & 8  & 3.7100 & 74.20 & 11 \\
        ie    & 4.7142 & 94.28 & 7  & 4.8500 & 97.00 & 5 \\
        inter & 3.3636 & 67.27 & 11 & 2.7750 & 55.50 & 10 \\
        site  & 2.5000 & 50.00 & 7  & 2.4642 & 49.28 & 7 \\
        soe   & 3.4285 & 68.57 & 7  & 3.3000 & 66.00 & 5 \\
        \bottomrule
    \end{tabular}
\end{table}

\begin{figure}[ht!]
    \centering
    \includegraphics[width=\textwidth]{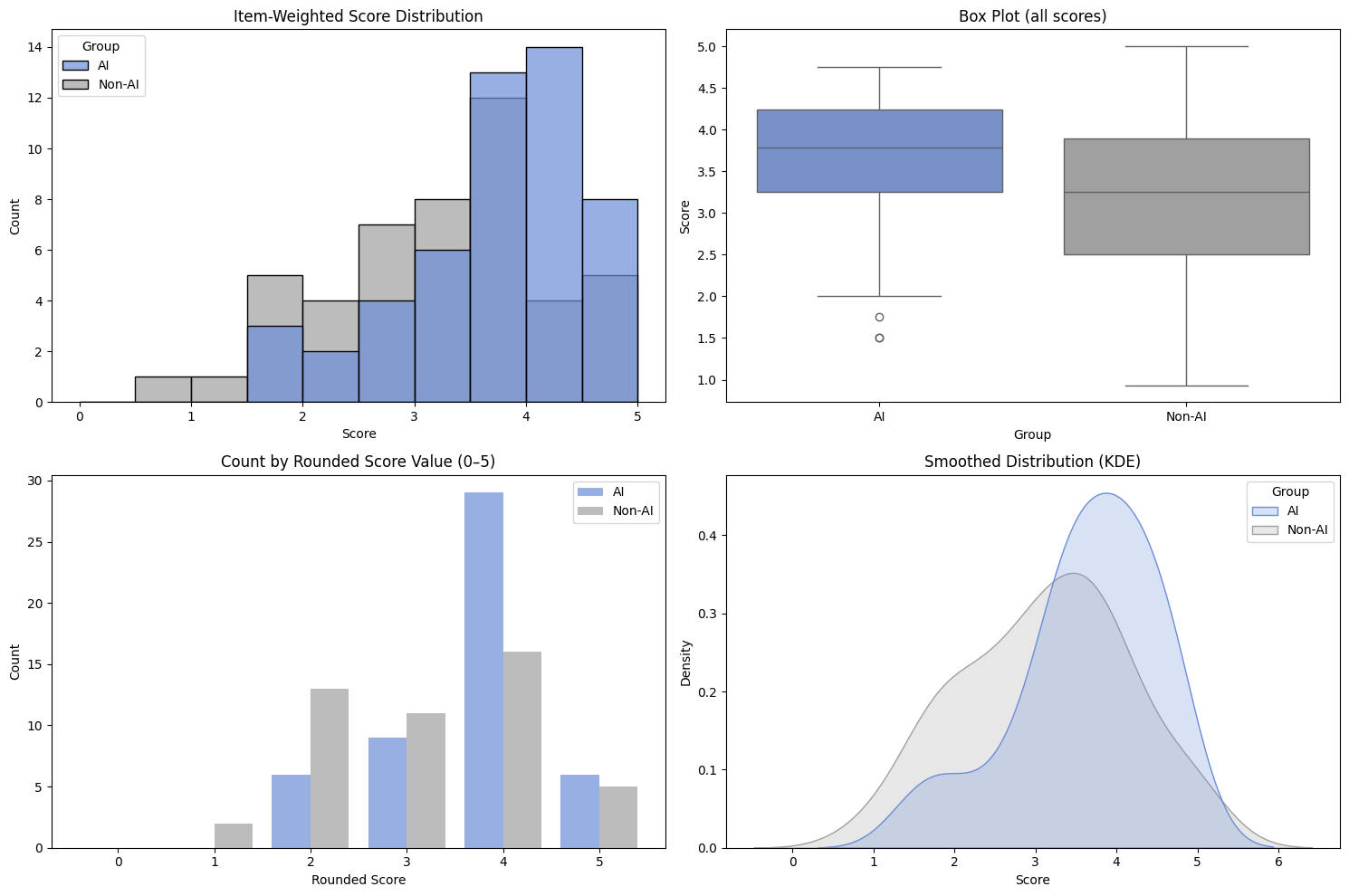}
    \caption{Distribution and comparison of item-weighted accuracy scores between AI-assisted and unassisted conditions. (A) Histogram showing count distribution of item-weighted scores, with AI-assisted scores (blue) and unassisted scores (gray) overlaid. (B) Box plot displaying median, quartiles, and range of scores for both conditions. (C) Count distribution by rounded score values on 0--5 scale, comparing frequencies between AI and non-AI groups. (D) Kernel density estimation curves showing smoothed probability distributions for both conditions.}
    \label{fig:accuracy_scores}
\end{figure}

We emphasize that the above comparison involves a RAG system with gpt-4o mini as the generation model, not the most recent configuration from Section~\ref{sec:rag_vs_standalone} which tends to produce more complete outputs. This, combined with the very high level of detail from the ground truth obtained through Section~\ref{sec:data_annotation_approach}, typically higher than human annotation on most categories, can explain the overall lower scores above relative to the best configurations from Section~\ref{sec:rag_vs_standalone}.

\subsubsection{User satisfaction and preferences}

Following completion of both the AI-assisted and manual extraction tasks, CRC participants are asked to complete a short survey assessing their perceptions of AI assistance. Twelve of the thirteen participants completed the survey.

Table~\ref{tab:user_satisfaction} summarizes the response distributions. Overall, the results demonstrate a strong perceived usefulness of the AI assistance, with 11 participants finding the task easier than manual extraction. Furthermore, the tool successfully addressed cognitive workload, with a majority of CRCs (11/12) reporting a reduction in the required thinking or decision-making. We also note high system adoption intent, as nearly all participants indicated they would likely choose AI assistance for future tasks.

\subsubsection{Other observations and examples}

Analysis of individual protocol abstractions reveals three distinct error patterns that contribute to reduced performance in unassisted tasks.

\paragraph{Empty fields}

In unassisted cases, participants frequently leave complex fields blank when information is challenging to locate within lengthy protocols. For example, when extracting safety monitoring plans, unassisted participants often provide no response, while AI-assisted participants extract comprehensive details:

\begin{lstlisting}[style=json]
# unassisted
{"safety_monitoring_plan": {}}
# ai assisted (same protocol)
{"safety_monitoring_plan": {
  "safety_monitoring_plan": "The safety of inhaled treprostinil will be evaluated by comparing parameters such as AEs, oxygenation, pulmonary function, clinical laboratory parameters, vital signs, 12-Lead ECG, hospitalizations due to cardiopulmonary indications, and exacerbations of underlying lung disease between the two treatment groups. All AEs will be assigned a MedDRA preferred term and system organ class for reporting purposes. The Data Monitoring Committee (DMC) will oversee safety monitoring, meeting regularly as outlined in the DMC charter.",
  "Safety monitoring procedures and frequency": "Interim analyses for safety data will be performed at the request of the DMC. The DMC will meet regularly to monitor the safety of the study, with meetings occurring as outlined in the DMC charter. Study monitors will periodically contact the site and conduct on-site visits to confirm ethical treatment of subjects, assess study progress, review data collected, conduct source document verification, verify drug accountability, and identify any issues requiring resolution."
}}
\end{lstlisting}

Another example of this phenomenon:
\begin{lstlisting}[style=json]
# unassisted
{"product_name": "Pembrolizumab", "dose": "200 mg", "administration": "every 3 weeks", "restrictions": ""}
# restriction is clearly missed here, however on ai assisted:
{"product_name": "Pembrolizumab", "dose": "200 mg", "administration": "IV infusion", "restrictions": "Stop infusion for Grade 3 reactions; no subsequent dosing if severe reactions occur."}
\end{lstlisting}

\paragraph{Non-informative responses:}

Unassisted tasks sometimes provide generic responses that fail to capture specific required information. In one case involving drug storage requirements, the response repeats identical boilerplate language (``Clinical supplies must be stored in a secure, limited-access location under the storage conditions specified on the label'') across three distinct fields (temperature, environmental conditions, and storage location). The AI-assisted version provides specific details: temperature requirements (2--8\textdegree C), environmental conditions (protect from light, do not shake), and storage specifications to the participant helping them to find the right answer.

\begin{lstlisting}[style=json]
# unassisted
{"product_name": "Pembrolizumab",
 "temperature_requirements": "Clinical supplies must be stored in a secure, limited-access location under the storage conditions specified on the label.",
 "environmental_conditions": "Clinical supplies must be stored in a secure, limited-access location under the storage conditions specified on the label.",
 "storage_location": "Clinical supplies must be stored in a secure, limited-access location under the storage conditions specified on the label."}
# AI-assisted, product code is added, specific temperature and storage location is added:
{"product_name": "Pembrolizumab (MK-3475)",
 "temperature_requirements": "2-8 C (36-46 F)",
 "environmental_conditions": "Store in original box to protect from light; do not shake",
 "storage_location": "Secure, limited-access location"}
\end{lstlisting}

\paragraph{Incorrect values}

Sometimes direct transcription errors occur in unassisted tasks, including incorrect product names (TPI 28 vs. TPI 287) and truncated NCT identifiers (NCT0218782 vs. NCT02187822), which could have significant implications for protocol accuracy.

\begin{lstlisting}[style=json]
# unassisted:
{"treatment_level_information": [{"product_name": "TPI 28", "...": "..."}]}
# AI-assisted:
{"treatment_level_information": [{"product_name": "TPI 287", "...": "..."}]}
...
# unassisted:
{"study_nct_id": "NCT0218782"}
# AI-assisted:
{"study_nct_id": "NCT02187822"}
\end{lstlisting}

\section{Discussion}\label{sec:discussion}

This work provides evidence that clinical-research-specific RAG methods, paired with a SoE-focused vision approach, can extract protocol information efficiently and with high quality across diverse therapeutic areas and document styles. In direct comparisons on 23 publicly available protocols, our specialized RAG reliably matched or exceeded standalone LLM prompting, with the largest gains observed in categories where relevant details are scattered across multiple sections (adverse events, interventions, site requirements) and in tabular SoE content.

Relative to prior extraction work, our contribution offers extended extraction coverage and differs in two practical ways on overlaps. For eligibility extraction, existing approaches (e.g. \cite{datta2024, liu2021, snorkel2022}) typically start from pre-extracted criteria lists that may not fully preserve protocol-faithful wording and context, whereas our pipeline treats the protocol document itself as the source of truth. Conversely, those works often provide finer downstream structuring of criteria, which we plan to incorporate in future iterations through dedicated LLM-based post-structuring. For SoE extraction, \cite{snorkel2022} extracts from full PDFs but focuses primarily on procedure lists while \cite{kramer2025} does not explicitly handle table-page and footnotes detection or certain layout complexities (e.g. images, cohorts). Our vision-enabled workflow processes page images, detects SoE table regions, separates multi-cohort structures, and produces outputs closer to operational use. Despite this broader scope, the performance of our Claude Opus~4.5 configuration seems comparable while being operationally simpler.

These findings extend and complement a growing body of work applying RAG and LLMs to clinical trial documents. Recent work shows that RAG substantially improves LLM performance for protocol section generation, raising clinical thinking and logic scores from approximately 40\% to 80\% \cite{markey2025}. Our work addresses the complementary and prerequisite challenge of protocol extraction: structured operational data must first be reliably extracted from existing protocols before generation or downstream automation can proceed. Similarly, RAG was applied to assess regulatory compliance of protocol statistical analysis plans \cite{waikar2026}, focusing on a narrower compliance checking task. In contrast, our system covers a broader operational spectrum (interventions, adverse events, SoE, site requirements, eligibility) and provides element-level chunk references to support auditability, extending the applicability of RAG in the clinical trial context.

In our controlled experiment with CRCs, AI assistance not only significantly reduces time-to-completion, but also increases accuracy, while being largely perceived as helpful and lowering cognitive load.

The efficiency and satisfaction gains observed in our controlled experiment are consistent with the broader pattern reported in the literature on LLM deployment in clinical workflows. A recent systematic review found that across real-world implementations, LLMs were associated with improvements in operational efficiency, user satisfaction, and reduced workload \cite{artsi2025}, while also noting that real-world adoption remains constrained by generalizability limitations and lack of post-deployment monitoring, challenges that apply equally here and motivate the real-world pilot deployments described in future work. Looking further ahead, researchers envision AI agents acting as orchestrators across the clinical trial lifecycle, coordinating activities by leveraging trial protocols and patient data dynamically \cite{badani2025}. The structured, machine-readable protocol abstractions produced by our system represent a foundational layer that such agentic frameworks would require, positioning protocol intelligence as an enabler of the broader AI-assisted trial operations.

Our findings suggest immediate practical value for clinical operations. Standardized representations of protocol content enable downstream automation and decision support for feasibility, start-up, site activation, and ongoing monitoring. By attaching transparent, chunk-level references to each extracted element, the approach supports auditability and human review, facilitating traceability and reproducibility. As institutions modernize clinical research infrastructure, similar upstream protocol intelligence could serve as a hub from which study information is propagated across systems (e.g. CTMS, feasibility tools, EDC), reducing manual effort and variability. Automation of routine structuring and integration of protocol content can free research teams from low-leverage administrative tasks, redirecting focus towards patient-centric care and scientific decision-making. Ultimately, efficiency and quality gains would reduce delays, accelerate evidence generation and improve patient outcomes by expanding access to high-quality, compliant research.

Important limitations remain, however, among them:
\begin{itemize}
    \item establishing a complete and accurate ground truth at the scale required for this paper represents a significant challenge. The density of medical information, spread of related information across sections, and subjectivity around which element components are essential or optional are such that even expert reviewers are prone to oversight and fatigue. Consequently, a small group of expert annotators cannot guarantee a truly exhaustive ground truth while increasing their headcount, even if it could improve completeness, is prohibitively labor-intensive and fundamentally unscalable. Furthermore, the level of verbosity of such ground truth may exceed what clinical practitioners actually require or extract in a real-world, operational context, potentially introducing a gap between annotation-based evaluation and real-world clinical metrics;
    \item our evaluation methodology, leveraging hybrid human--AI ground truth generation and LLM-as-judge with human quality control, prioritizes scalability but may inherit model biases. Real-world validation against expert adjudication (e.g. data edits and validation from clinical experts) better reflecting real-world conditions and scale in a clinical context is required to corroborate true performance;
    \item results may not fully generalize to certain trials, complex adaptive designs, or highly specialized domains without further tailored schemas and prompts,
    \item our dataset, while spanning public protocols in multiple therapeutic areas with heterogeneous formatting and variable quality (see Appendix~\ref{sec:protocol_selection_annex}), is modest and cannot capture the full spectrum of real-world variability. The performance metrics reported represent baseline feasibility and relative comparisons. Sustained performance on out-of-distribution protocols in realistic clinical settings (e.g. with evolving document styles and standards) will require ongoing validation and adaptation.
\end{itemize}

Future work should include real-world pilot deployments at clinical sites, data element confidence/risk scoring (more generally, robust performance and safety monitoring), expansion of semi-structured schemas to cover further operational needs, deeper alignment with emerging data standards (e.g. CDISC USDM), integration with existing site systems (e.g. CTMS, IRB platforms), protocol amendment handling, RAG pipeline hyperparameter and prompt tuning.

Governance and safety are essential for clinical site adoption. For that purpose, our foundational designs and methods ensure privacy preservation across all data handling stages, maintain comprehensive audit trails and compliance with healthcare standards, and involve a complementary process providing continuous automated error monitoring and inspection in addition to embedded human approval at key checkpoints to mitigate hallucinations and propagation of AI-generated errors. Clear user roles and responsibilities provided through proper training (e.g. what is expected from the automation and expert judgment) also help teams adopt AI assistance safely and confidently.

\section{Conclusion}\label{sec:conclusions}

The quantified improvements measured in this work (higher extraction accuracy, reduced completion times and strong user preference) support the practical value of integrating clinical-trial-specific RAG-based protocol extraction into research workflows. By combining targeted retrieval, transparent referencing, and structured outputs with expert review, research teams can scale protocol comprehension and operational initiation of trials while maintaining rigor and accountability. With responsible deployment and continued validation, these methods show promise in helping to address the growing complexity of clinical trials and unlocking further efficiencies across the trial lifecycle.

\section*{Data availability}
Data will be made available on request.

\section*{Ethics statement}
All procedures involving human participants were conducted in accordance with national research ethical standards, privacy rights of human subjects have been observed and informed consent was obtained for experimentation with human subjects.

\section*{AI declaration}
During the preparation of this work the authors used Google Gemini Enterprise and GitHub Copilot Enterprise in order to assist linguistic review and enhancement. After using this tool/service, the authors reviewed and edited the content as needed and take full responsibility for the content of the published article.

\section*{Funding}
Authors report financial support provided by Banting Health AI Inc. as well as an employment relationship with Banting Health AI Inc.

\section*{Declaration of competing interest}
Francois Charest reports financial support was provided by Banting Health AI Inc. Madison Wright reports financial support was provided by Banting Health AI Inc. Ramtin Babaeipour reports financial support was provided by Banting Health AI Inc. Francois Charest reports a relationship with Banting Health AI Inc. that includes: employment and equity or stocks. Madison Wright reports a relationship with Banting Health AI Inc. that includes: employment and equity or stocks. Ramtin Babaeipour reports a relationship with Banting Health AI Inc. that includes: employment and equity or stocks. If there are other authors, they declare that they have no known competing financial interests or personal relationships that could have appeared to influence the work reported in this paper.

\bibliographystyle{elsarticle-num}
\bibliography{references}

\clearpage

\appendix
\section{Appendices}\label{sec:annexes}

\begin{samepage}
\subsection{Protocol documents selection}\label{sec:protocol_selection_annex}

Starting from identifiable protocol documents (using the ``Study Documents'' field), the following filtering logic is applied:

\begin{lstlisting}[style=json, caption={Filtering logic for studies.}, label={lst:filtering}]

studies_filtered = studies[
    (studies["Interventions"].str.startswith("intervention_type:Drug"))
    & (studies["Study_Design_Primary_Purpose"] == "Treatment")
    & (studies["Location_Countries"].str.contains("Canada|United States"))
]

studies_filtered["therapeutic_area"] = "other"

studies_filtered.loc[
    studies_filtered["Primary_Disease_or_Condition"]
    .str.lower()
    .str.contains(
        "cancer|carcinoma|neoplasm|myeloma|lymphoma|leukemia|tumor
        |melanoma|neuroblastoma|glioblastoma|sarcoma|atrial",
        case=False,
    ),
    "therapeutic_area",
] = "oncology"

studies_filtered.loc[
    studies_filtered["Primary_Disease_or_Condition"]
    .str.lower()
    .str.contains(
        "heart|cardiac|cardial|stroke|artery|coronary|cardio|hypertension",
        case=False,
    ),
    "therapeutic_area",
] = "cardiology"

\end{lstlisting}
\end{samepage}

As a last step, we filter out PDF documents that appear scanned (e.g. using a heuristic based on the number of pages and file size) to ensure that the final dataset contains only text-based protocols that can be processed by the RAG system.

Note that the above filtering intentionally excludes timestamps, other conditions on formatting or document quality. This combined with uniform random selection in each category allows us to obtain a dataset that represents, despite its modest size, significant diversity in publicly available protocols. This is important to evaluate the robustness of the RAG system across different document styles, formatting and quality levels.

In particular, the selected protocols span diverse sponsors (including investigator-led studies), phases, publication eras (pre-2010 to post-2020), and overall structures, with heterogeneous schedule-of-events (tabular, list-based, and narrative) and eligibility/endpoint formats, HTML-derived prints (rendered as table-like structures), and varying types of diagrams and amendment histories.

\subsection{Extracted semi-structured data models}

\begin{lstlisting}[style=json, caption={Example json output for intervention/treatment information.}, label={lst:intervention}]
{
    "arms": [
        {
            "arm_name": "Active PMA",
            "interventions": [
                {
                    "type": "Immunotherapy",
                    "name": "Proprietary Mixture of FDA Approved Allergens (PMA)",
                    "dosage": "0.15-0.60 ml",
                    "schedule": "Twice-weekly clinic visits for 8 weeks"
                }
            ]
        },
        {
            "arm_name": "Placebo",
            "interventions": [
                {
                    "type": "Placebo",
                    "name": "Saline Placebo",
                    "dosage": "0.15-0.60 ml",
                    "schedule": "Twice-weekly clinic visits for 8 weeks"
                }
            ]
        }
    ],
    "treatment_level_information": [
        {
            "product_name": "PMA (Proprietary Antigen Preparation)",
            "dose": "The dosage (dilution) of PMA will be the same for all subjects and will begin with vial number #5 [...].",
            "administration": "Sub-cutaneous administration using [...].",
            "restrictions": "All patients are required to remain in the clinic under close observation by the study nurse for a period of 30 minutes following their injection [...].",
            "modifications": "Subjects that develop local swelling greater than 50mm will have their next dose decreased by one step [...]."
        }
    ]
}
\end{lstlisting}

\begin{lstlisting}[style=json, caption={Example json output for general information and endpoints.}, label={lst:gen_info}]
{
    "general_information": {
        "study_nct_id": "NCT03360071",
        "protocol_id": "PMA 001",
        "protocol_version": "6.9.1",
        "protocol_date": "2018-08-07",
        "study_title": "A double-blind, prospective, parallel group evaluation of a novel biologic therapy for perennial allergic rhinitis",
        "sponsor_name": "Relez Therapeutics, LLC",
        "phase": "Phase 2",
        "therapeutic_area": "allergology and immunology",
        "disease_or_condition": "perennial allergic rhinitis",
        "allocation": "Randomized",
        "masking": "Double Masking",
        "number_of_sites": 1,
        "estimated_duration": "14 weeks"
    },
    "objectives_and_endpoints": {
        "primary_objectives": [
            "To evaluate whether there is a difference in the signs and symptoms of moderate to severe allergic rhinitis when treated with a proprietary mixed preparation of FDA approved allergens [...]"
        ],
        "secondary_objectives": [
            "To quantify allergy medication use and adverse events.",
            "Serum will also be stored to evaluate changes in IgE and IgG subclasses.",
            "..."
        ],
        "primary_endpoints": [
            "Quantitative comparison of clinical response of active PMA immunotherapy compared to placebo",
            "Total combined symptom and medication score (DCS)",
            "Total Nasal Symptoms Score (TNSS)"
        ],
        "secondary_endpoints": ["..."],
        "measurement_method": [
            "Validated instruments that assess study subject clinical response, physician global assessment, and medication use."
        ],
        "endpoint_timepoint": "8 weeks"
    }
}
\end{lstlisting}

\begin{lstlisting}[style=json, caption={Example json output for schedule of events.}, label={lst:soe}]
{
    "schedule_of_events": [
        {
            "visit_number": 1,
            "visit_time": "Week 1",
            "procedures": [
                {"procedure_name": "Discussion and Recruitment"},
                {"procedure_name": "Informed consent"},
                {"procedure_name": "..."},
                {"procedure_name": "Blood Draw"}
            ]
        },
        {
            "visit_number": 2,
            "visit_time": "Week 3 to 11",
            "procedures": [
                {"procedure_name": "Administer immunotherapy/safety f/u"},
                {"procedure_name": "..."},
                {"procedure_name": "AE assessment"}
            ]
        },
        {
            "visit_number": 18,
            "visit_time": "Week 14",
            "procedures": [
                {"procedure_name": "Blood Draw"},
                {"procedure_name": "..."},
                {"procedure_name": "AE assessment"}
            ]
        }
    ]
}
\end{lstlisting}

\begin{lstlisting}[style=json, caption={Example json output for inclusion/exclusion criteria.}, label={lst:ie}]
{
    "inclusion": [
        "The study subject is currently experiencing active symptoms and signs of moderate to severe PAR with or without SAR component defined as score of at least 28 on the RQLQ and 6 on the TNSS scale. Scores will be determined on the first and second visit (after the 14 day medication washout period). Both scores exceeding their threshold on either day will qualify.",
        "Skin test strongly positive (wheal at least 5mm diameter and 2mm greater than negative control) to at least 6 of 48 allergy skin prick tests including at least one species of dust mite.",
        "The subject is a male or a non-pregnant, non-lactating female between the ages of 18 and 65."
    ],
    "exclusion": [
        "The subject has received any investigational compound within 30 days prior to screening.",
        "...",
        "Subject has a history of cancer, other than squamous cell or basal cell carcinoma of the skin that has not been in full remission for at least 5 years prior to Screening. (A history of treated CIN I, II, or CIN III [cervical intraepithelial neoplasia] is allowed.)"
    ]
}
\end{lstlisting}

\begin{lstlisting}[style=json, caption={Example json output for adverse event definitions.}, label={lst:ae_definitions}]
{
  "ae_definitions_and_classifications": {
    "ae_definition": "An AE is defined as any untoward medical occurrence in a clinical investigation subject administered a drug; [...]",
    "sae_definition": "An SAE is defined as any untoward medical occurrence that at any dose results in death, is life-threatening, requires inpatient hospitalization or prolongation of existing hospitalization, [...].",
    "severity_grading": "Mild: The event is transient and easily tolerated by the subject. Moderate: [...]",
    "ae_relationship": "Related: An AE that follows a reasonable temporal sequence from administration of the immunotherapy. Not Related: [...]."
  },
  "reporting_requirements": {
    "reporting_timeframes": "A sponsor SAE form must be completed and signed by the investigator immediately or within 24 hours of first onset or notification of the event.",
    "ae_data_collection_requirements": "Event term, start and stop date, severity, [...].",
    "reporting_contacts": "***"
  },
  "safety_monitoring_and_management": {
    "safety_monitoring_plan": "All subjects experiencing AEs must be monitored until the symptoms subside [...].",
    "discontinuation_criteria": "The subject has experienced a PTE or AE that requires early termination because [...].",
    "emergency_procedures": ""
  },
  "specific_ae_information": {
    "expected_aes": "Typical but uncommon side effects have included mild-moderate local swelling.",
    "potential_risks": "The most common side effect from allergen immunotherapy [...] There is a very small chance of having a serious allergic reaction.",
    "concomitant_medication_restrictions": "Immunotherapy that is not part of the investigation, [...].",
    "special_population_considerations": "If female, the subject is pregnant or lactating or intending to become pregnant before [...]."
  }
}
\end{lstlisting}

\begin{lstlisting}[style=json, caption={Example json output for site requirements.}, label={lst:site_req}]
{
  "site_requirements": [
    {
      "equipment": ["refrigerator", "sterile syringe", "27G 1/2 inch syringe", "temperature log"],
      "certifications": ["GCP certification"],
      "sample_handling": {
        "sample_description": "Blood samples collected for immunoglobulin level measurement, including specific IgE, IgG4, and possibly IgG3.",
        "processing_timeline": "Blood will be drawn at the start and completion of the study. Serum collected at the screening visit will be stored for later analysis. Blood will be redrawn at the final visit (week 14).",
        "storage_conditions": "refrigerated at 4 degrees Celsius",
        "transport_requirements": "All samples will be collected in accordance with acceptable laboratory procedures.",
        "access_control": "The investigator or designee is responsible for transcribing or attaching laboratory results to the CRF. The investigator will maintain a copy of the laboratory accreditation and the reference ranges for the laboratory used.",
        "chain_of_custody": "Maintain current records of the receipt, administration, and disposition of sponsor-supplied immunotherapy, and return all unused sponsor-supplied study solutions to the sponsor."
      }
    }
  ],
  "ip_storage": [
    {
      "product_name": "Investigational Immunotherapy Solutions and Placebo Saline Solutions",
      "temperature_requirements": "Refrigerated at 4 degrees Celsius",
      "environmental_conditions": "Stored in an appropriate, limited-access, secure place",
      "storage_location": "Original container until dispensed",
      "access_restrictions": "Limited-access, secure place",
      "accountability_tracking": "A daily temperature log of the drug storage area must be maintained every working day"
    }
  ]
}
\end{lstlisting}

\begin{table}[H]
    \centering
    \small
    \caption{Element weights by data element category for weighted evaluation scoring. Weights within each category sum to 1.0.}
    \label{tab:component_weights}
    \begin{tabular}{p{2cm}p{7cm}r}
        \toprule
        \textbf{Category} & \textbf{Element} & \textbf{Weight} \\
        \midrule
        gen & Study NCT ID & 0.15 \\
        gen & Study Title & 0.08 \\
        gen & Protocol ID, Version, Date & 0.07 \\
        gen & Sponsor Name & 0.05 \\
        gen & Phase & 0.10 \\
        gen & Therapeutic Area & 0.01 \\
        gen & Disease/Condition & 0.03 \\
        gen & Allocation & 0.02 \\
        gen & Masking & 0.02 \\
        gen & Target Enrollment & 0.00 \\
        gen & Countries and Sites & 0.00 \\
        gen & Timeline (Start/End/Duration) & 0.00 \\
        gen & Primary Objectives & 0.25 \\
        gen & Secondary Objectives & 0.02 \\
        gen & Primary/Secondary Endpoints & 0.20 \\
        \midrule
        ie & Inclusion Criteria & 0.50 \\
        ie & Exclusion Criteria & 0.50 \\
        ie & Demographics & 0.00 \\
        ie & Washout Period & 0.00 \\
        \midrule
        ae & AE Definition & 0.15 \\
        ae & SAE Definition & 0.20 \\
        ae & Severity Grading & 0.05 \\
        ae & AE Relationship Assessment & 0.02 \\
        ae & Reporting Timeframes & 0.12 \\
        ae & Data Collection Requirements & 0.08 \\
        ae & Reporting Contacts & 0.02 \\
        ae & Safety Monitoring Plan & 0.08 \\
        ae & Discontinuation Criteria & 0.12 \\
        ae & Emergency Procedures & 0.05 \\
        ae & Expected AEs & 0.04 \\
        ae & Potential Risks & 0.03 \\
        ae & Concomitant Medication Restrictions & 0.03 \\
        ae & Special Population Considerations & 0.01 \\
        \midrule
        inter & Intervention Arms & 0.50 \\
        inter & Treatment-Level Information & 0.50 \\
        \midrule
        site & Equipment & 0.20 \\
        site & Certifications & 0.10 \\
        site & Sample Handling & 0.35 \\
        site & IP Storage & 0.35 \\
        \midrule
        soe & Schedule of Events & 1.00 \\
        \bottomrule
    \end{tabular}
\end{table}

\begin{lstlisting}[style=json, caption={Example comparison showing subtle data variation.}, label={lst:subtle_variation}]
# AI output:
"exclusion": [
    "The subject has received any investigational compound within 30 days prior to screening.",
    "OR",
    "The subject has received allergen immunotherapy or SLIT in a previous clinical study or as a therapeutic agent within the past two years.",
    # p.13
]
# annotation:
"exclusion": [
    "The subject has received any investigational compound within 30 days prior to screening.",
    "OR",
    "The subject has received multi antigen immunotherapy or SLIT in a previous clinical study or as a therapeutic agent within the past two years.",
    # p.26
]
\end{lstlisting}

\subsection{Evaluation examples}\label{sec:evaluation_examples_annex}

\begin{lstlisting}[style=json, caption={Example comparison of adverse event definition information.}, label={lst:ae_comp}]
# Ground truth:
{"ae_definition": "An AE is defined as any untoward medical occurrence in a clinical investigation subject administered a drug; it does not necessarily have to have a causal relationship with this treatment."}

# AI output:
{"ae_definition": "An AE is defined as any untoward medical occurrence in a clinical investigation subject administered a drug; it does not necessarily have to have a causal relationship with this treatment. An AE can therefore be any unfavorable and unintended sign (e.g., a clinically significant abnormal laboratory finding), symptom, or disease temporally associated with the use of a drug whether or not it is considered related to the drug."}

# Evaluation output:
{"score": 5,
 "rationale": "The submitted json recovers 100% of the ground truth json to a human reader. The definition of an AE in the submitted json includes the exact wording of the ground truth json and adds additional information that is consistent with the ground truth. The extra information provided does not contradict the ground truth and is relevant to the context of defining an AE. Therefore, the meaning is preserved, and the additional details enhance understanding without introducing any discrepancies."}
\end{lstlisting}

\begin{lstlisting}[style=json, caption={Example comparison of disease or condition information.}, label={lst:disease_comp}]
# Ground truth:
{"disease_or_condition": "perennial allergic rhinitis"}

# AI output:
{"disease_or_condition": "allergic rhinitis"}

# Evaluation output:
{"score": 4,
 "rationale": "The submitted json specifies \"allergic rhinitis\" while the ground truth specifies \"perennial allergic rhinitis\". The term \"perennial\" indicates that the allergic rhinitis is year-round, as opposed to seasonal. While \"allergic rhinitis\" is a broader category that includes both perennial and seasonal types, the omission of \"perennial\" means that the submitted json does not fully capture the specific condition described in the ground truth. However, the core condition of \"allergic rhinitis\" is still present, which means the submitted json recovers a significant portion of the ground truth information, but not completely. Therefore, it scores a 4, as it recovers at least 80% of the ground truth information."}
\end{lstlisting}

\begin{lstlisting}[style=json, caption={Example comparison of country information.}, label={lst:countries_comp}]
# Ground truth:
{"countries": [{"country_name": "USA", "number_of_sites": 1}],
 "number_of_sites": 1}

# AI output:
{"countries": [],
 "number_of_sites": 1}

# Evaluation output:
{"score": 2,
 "rationale": "The submitted json captures the 'number_of_sites' correctly, which matches the ground truth. However, it fails to include the 'countries' information, which is a significant part of the ground truth json. The ground truth specifies that there is one site in the USA, but the submitted json has an empty list for 'countries'. This omission means that the submitted json recovers less than 40% of the ground truth information, as it lacks the geographical context provided by the 'countries' field. Therefore, the score is 2."}
\end{lstlisting}

\subsubsection{User satisfaction and preferences}\label{sec:user_satisfaction_annex}

The survey questions focus on three primary user experience constructs: Perceived Usefulness, Cognitive Load, and System Adoption Intent. These constructs were selected because reducing the high mental burden and time intensity of data extraction is critical for CRC retention and data quality.

To ensure an established methodological basis, items assessing mental demand and effort were informed by the NASA Task Load Index (NASA-TLX), a multidimensional scale for estimating subjective workload. Specifically, we attempted to measure ``Mental Demand'' and ``Effort'' to quantify the reduction in decision-making complexity provided by AI. Items evaluating satisfaction and the likelihood of future use were informed by the Technology Acceptance Model (TAM), specifically focusing on Perceived Usefulness (PU) (e.g., ``How helpful did you find the AI assistance?'') and Perceived Ease of Use (PEOU) (e.g., ``Did AI assistance make the task easier or harder?''). This approach allows for a more standardized comparison between the manual and AI-assisted workflows.

\begin{table}[ht!]
\centering
\caption{User experience survey results categorized by theoretical constructs ($N = 12$).}
\label{tab:user_satisfaction}
\begin{tabularx}{\linewidth}{>{\raggedright\arraybackslash}X >{\raggedright\arraybackslash}p{6cm}}
\toprule
\textbf{Construct \& Question} & \textbf{Response distribution} \\
\midrule
\multicolumn{2}{l}{\textit{\textbf{Perceived usefulness}}} \\
\midrule
How helpful did you find the AI assistance? &
\begin{tabular}[t]{@{}l@{}}
Extremely helpful (4)\\
Very helpful (4)\\
Somewhat helpful (3)\\
Not very helpful (1)\\
Not at all helpful (0)
\end{tabular} \\
\addlinespace
Did AI assistance make the task easier or harder compared to manual extraction? &
\begin{tabular}[t]{@{}l@{}}
Much easier (8)\\
Slightly easier (3)\\
About the same (1)\\
Slightly harder (0)\\
Much harder (0)
\end{tabular} \\

\midrule
\multicolumn{2}{l}{\textit{\textbf{Cognitive load}}} \\
\midrule
Did AI assistance reduce the amount of thinking or decision-making required? &
\begin{tabular}[t]{@{}l@{}}
Yes, significantly (5)\\
Yes, somewhat (6)\\
Not really (0)\\
Not at all (1)\\
It made it more difficult (0)
\end{tabular} \\

\midrule
\multicolumn{2}{l}{\textit{\textbf{System adoption intent}}} \\
\midrule
How satisfied are you with AI assistance compared to manual extraction? &
\begin{tabular}[t]{@{}l@{}}
Much more satisfied (8)\\
Slightly more satisfied (3)\\
About the same (1)\\
Slightly less satisfied (0)\\
Much less satisfied (0)
\end{tabular} \\
\addlinespace
How likely would you be to choose AI assistance again for similar tasks? &
\begin{tabular}[t]{@{}l@{}}
Very likely (5)\\
Likely (5)\\
Neutral (2)\\
Unlikely (0)\\
Very unlikely (0)
\end{tabular} \\
\bottomrule
\end{tabularx}
\end{table}

\end{document}